\documentclass[12pt,fullsize]{article}

\usepackage{amsfonts,amsmath,amssymb}
\usepackage{epsfig}

\newcommand{\ket}[1]{| #1 \rangle}
\newcommand{\bra}[1]{\langle #1 |}

\newcommand{\abs}[1]{| #1 |}

\renewcommand{\phi}{\varphi}

\begin{document}
\title{\textbf{Quantum Networks for Concentrating Entanglement}}
\date{November 2000}

\author{Phillip Kaye \thanks{prkaye@cacr.math.uwaterloo.ca,
Department of Combinatorics \& Optimization, University of
Waterloo, Waterloo, ON, N2T 2L1, Canada.} and Michele Mosca
\thanks{mmosca@cacr.math.uwaterloo.ca, Department of Combinatorics \& Optimization, University of
Waterloo.
}}

 \maketitle

\begin{abstract}
If two parties, Alice and Bob, share some number, $n$, of partially
entangled pairs of qubits, then it is possible for them to
concentrate these pairs into some smaller number of maximally
entangled states.  We present a simplified version of the algorithm
for such \emph{entanglement concentration}, and we describe
efficient networks for implementing these operations.
\end{abstract}
\section{Introduction}
The state of a single pure quantum bit, or \emph{qubit}, is
described by a vector in a 2-dimensional Hilbert space spanned by
basis vectors $\ket{0}$ and $\ket{1}$.  The state of $n$ pure
qubits (i.e. an $n$-qubit register) is described by a vector in a
$2^{n}$-dimensional Hilbert space which is the tensor product of
the 2-dimensional spaces for the states of each of the $n$ qubits.
Consider a 2-qubit register in a state described by the vector
$\ket{\Psi}=\frac{1}{\sqrt{2}}\ket{00}+\frac{1}{\sqrt{2}}\ket{11}$.
We call a pair of particles in this state an EPR pair, named after
Einstein, Podolsky and Rosen, who discussed such particle pairs in
their 1935 paper \cite{EPR35}.  It can easily be shown that this
vector cannot be factored into a tensor product of two 1-qubit
states. That is
\[ \ket{\Psi}=\frac{1}{\sqrt{2}}\ket{00}+\frac{1}{\sqrt{2}}\ket{11}\neq
(a_0\ket{0}+a_1\ket{1})\otimes(b_0\ket{0}+b_1\ket{1}) \] for any
$a_0, a_1, b_0, b_1$. The amount of entanglement present in a
bipartite quantum system can be quantified, and for this purpose
we will treat a single EPR pair as possessing one unit of
entanglement.

In many scenarios involving quantum communication, an essential
ingredient is the sharing of an EPR pair by Alice (the `sender' of
some information) and Bob (the `receiver').  For example, when
Alice and Bob share an EPR pair, they are able to perform quantum
teleportation, a process useful for communicating quantum
information. Using protocols involving the sharing of EPR pairs,
some distributed computation tasks can be achieved using fewer
bits than could be achieved using only a classical channel (see
e.g. \cite{BCW98} and \cite{Raz99}).

Suppose Alice and Bob share a known entangled pair of qubits
\[ \ket{\Psi} = \alpha_{00}\ket{0}\ket{0} +
                        \alpha_{01}\ket{0}\ket{1} +
                         \alpha_{10}\ket{1}\ket{0} +
                        \alpha_{11}\ket{1}\ket{1}  ,\]
                        where the first qubit is in Alice's possession
                        and the second qubit in Bob's.
The Schmidt decomposition for this bipartite system allows us to
express the state of this pair of qubits as
\[ \ket{\Psi} = \alpha \ket{a_0} \ket{b_0} + \beta \ket{a_1} \ket{b_1} ,\]
for some non-zero positive real numbers $\alpha$ and $\beta$, and
unit vectors $\ket{a_0}$ and  $\ket{a_1}$ that form a basis for
Alice's system, and unit vectors $\ket{b_0}$ and $\ket{b_1}$ that
form a basis for Bob's system. Since Alice and Bob can each
locally perform the one-qubit unitary operations
\[  \ket{a_0} \rightarrow \ket{0}, \ket{a_1} \rightarrow \ket{1} \]
and
\[  \ket{b_0} \rightarrow \ket{0}, \ket{b_1} \rightarrow \ket{1} \]
respectively, we will assume that Alice and Bob share an entangled
state of the form
\[ \alpha \ket{00} +\beta \ket{11} .\]

If $\abs{\alpha}=\abs{\beta}=\frac{1}{\sqrt{2}}$, then the state is
an EPR state, and is said to be \emph{maximally entangled}.  If
$\abs{\alpha}\neq\abs{\beta}$ then the state is less entangled, and
if either $\abs{\alpha}$ or $\abs{\beta}$ equal 0, then the state
is completely non-entangled.

Consider a $2n$-qubit system of the form $\ket{\Psi} =
\left(\alpha\ket{00}+\beta\ket{11}\right)^n$, shared by two
parties, Alice and Bob, where $\abs{\alpha}\neq\abs{\beta}$.  Now
suppose Alice and Bob want to share some maximally entangled EPR
pairs for some communication task.  A natural question is ``how
many EPR pairs can Alice and Bob distill out of $\ket{\Psi}$,
performing local operations and communicating classically''?  An
upper bound on the expected number of EPR pairs that can be
distilled is the ``entropy of entanglement'' of $\ket{\Psi}$
defined to be von Neumann entropy of either $\rho_A = \mbox{Tr}_B
\ket{\Psi}\bra{\Psi}$ or $\rho_B = \mbox{Tr}_A
\ket{\Psi}\bra{\Psi}$.  These quantities are both equal to the
Shannon entropy of the eigenvalues of $(\ket{\Psi}\bra{\Psi})^{n}$
(which are the squares of Schmidt coefficients of the state
$\ket{\Psi}^{n}$). This quantity equals $n$ times the von Neumann
entropy of $\ket{\Psi}\bra{\Psi}$, namely $n H(|\alpha^2|)$, where
\mbox{$H(p)=p\log\left(\frac{1}{p}\right)+(1-p)\log\left(\frac{1}{-1p}\right)$}.
For example, the Von Neumann entropy of an EPR pair is
$H(|(\frac{1}{\sqrt{2}})^2|) = 1$.

The process of distilling EPR pairs out of $\ket{\Psi}$ is called
\emph{entanglement concentration}. Local operations for performing
entanglement concentration have been by Bennett, Bernstein,
Popescu and Schumacher in \cite{BBPS95}. The expected amount of
concentrated entropy of entanglement is
\begin{equation} \label{equation1}
 \sum_{j=1}^{n-1} |\alpha^2|^{n-j} (1-|\alpha^2|)^j {n \choose j} \log_2 {n \choose j}
 \end{equation}
and they show that this quantity is in $n H(|\alpha^2|)
- O(\log n)$.

In section \ref{operations} we describe the approach detailed in
\cite{BBPS95}. We then describe a new way of extracting a specific
number of EPR pairs instead of the method suggested in
\cite{BBPS95}. In section \ref{network} we will give a description
of a quantum network for performing the main local basis change
necessary for performing entanglement concentration. In section
\ref{summary} we summarise how to implement entanglement
concentration.

\section{Local Operations for Entanglement
Concentration} \label{operations}

 Consider concentrating the entanglement of the
state $\ket{\Psi}$ defined above, and without loss of generality
we assume that $\alpha, \beta$ are positive real numbers. Consider
the case for $n=3$ qubits:
\begin{equation*}
\begin{split}
\left(\alpha \ket{00}+\beta \ket{11} \right)^3 \hspace{1mm} =
\hspace{1mm} & \alpha^3 \ket{\overset{alice}{0} \hspace{1mm}
\overset{bob}{0}} \ket{ \overset{alice}{0} \hspace{1mm}
\overset{bob}{0}} \ket{\overset{alice}{0} \hspace{1mm}
\overset{bob}{0}}\\ + & \alpha^2 \beta \left(\ket{00}\ket{00}\ket{11} +
 \ket{00} \ket{11} \ket{00} + \ket{11} \ket{00} \ket{00} \right)  \\ +
 & \alpha \beta^2 \left(\ket{00} \ket{11} \ket{11} +
 \ket{11} \ket{00} \ket{11} + \ket{11} \ket{11} \ket{00} \right)  \\ +
 & \beta^3 \ket{11} \ket{11} \ket{11}.
\end{split}
\end{equation*}

Separating Alice's qubits from Bob's, we can re-write the above
state as:
\begin{equation*}
\begin{split}
\alpha^3 \overset{alice} {\ket{000}} \hspace{1mm}
\overset{bob}{\ket{000}} + \alpha^2\beta \left( \ket{001}\ket{001}
+ \ket{010}\ket{010} + \ket{100}\ket{100}\right) \\ +
\alpha\beta^2 \left( \ket{011}\ket{011} + \ket{101}\ket{101} +
\ket{100}\ket{100}\right) + \beta^3 \ket{111}\ket{111}.
\end{split}
\end{equation*}

\begin{figure}[h]
\begin{center}
\epsfig{file=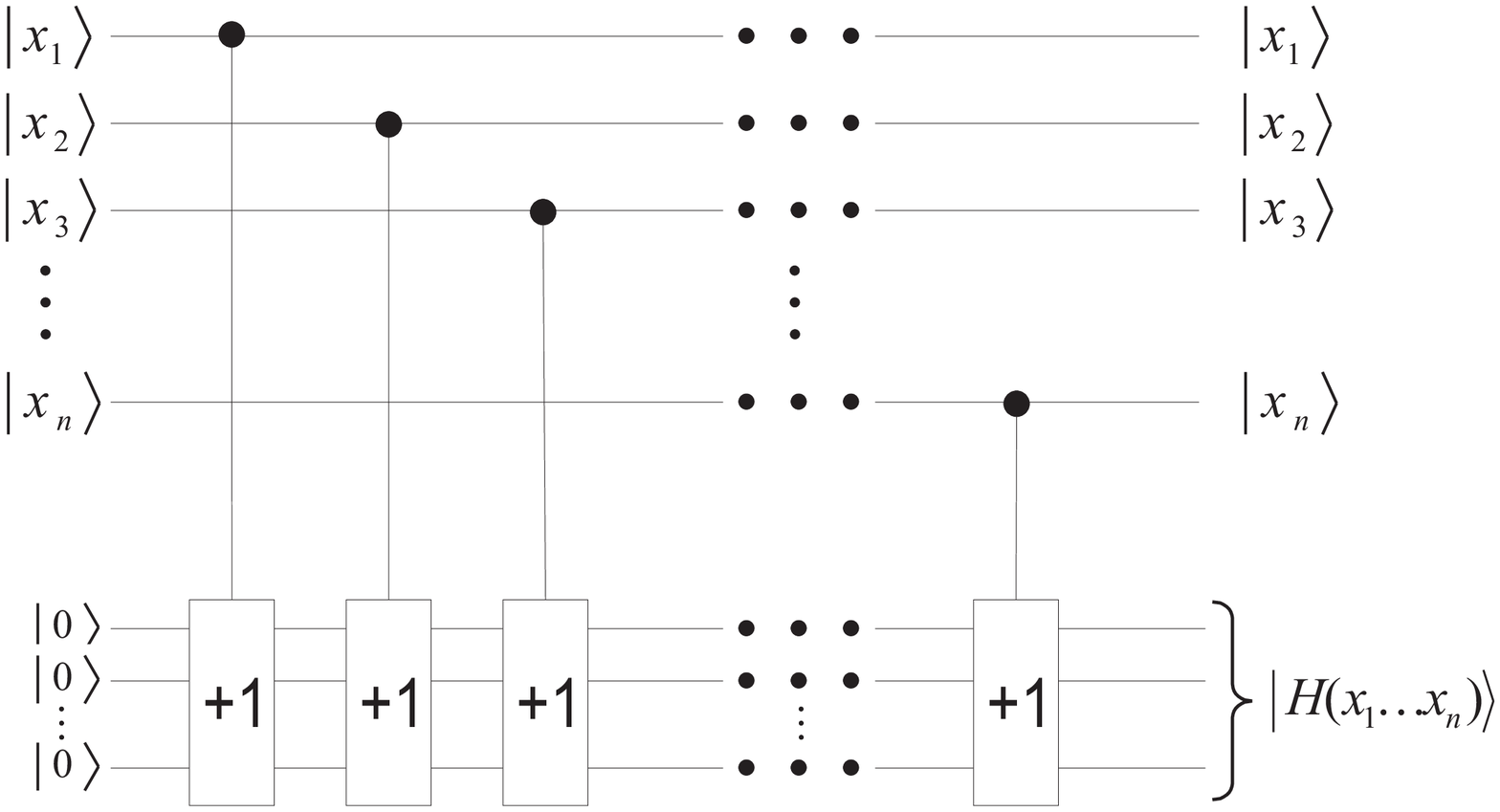,height=6cm,width=10cm,clip=}
\caption{Network to compute the Hamming weight. $\ket{x_1\cdots
x_n}\ket{00\cdots0}\longrightarrow\ket{x_1\cdots
x_n}\ket{\mbox{H}(x_1\cdots x_n)}$.} \label{hwfig}
\end{center}
\end{figure}

In general, if we have $n$ copies of
$\alpha\ket{00}+\beta\ket{11}$, by appropriately reordering the
qubits we get:
\begin{equation*}
\begin{split}
\hspace{1mm} \alpha^n\overset{a}{\ket{\mathbf{0}}}
\overset{b}{\ket{\mathbf{0}}} + & \alpha^{n-1}\beta\left(
\sum_{\mbox{H}(\mathbf{x})=1}\overset{a}{\ket{\bf
x}}\overset{b}{\ket{\bf x}}\right) +
\\&\alpha^{n-2}\beta^2\left(\sum_{\mbox{H}(\mathbf{x})=2}\overset{a}{\ket{\mathbf{x}}}
\overset{b}{\ket{\mathbf{x}}}\right) + \ldots + \\ &
\alpha\beta^{n-1}\left(\sum_{\mbox{H}(\mathbf{x})=n-1}\overset{a}{\ket{\mathbf{x}}}
\overset{b}{\ket{\mathbf{x}}}\right) +
\beta^n\overset{a}{\ket{\mathbf{1}}} \overset{b}{\ket{\mathbf{1}}}
\\ = & \sum_{j=0}^{n}\alpha^{n-j}\beta^j\left(\sum_{\mbox{H}(\mathbf{
x})=j}\overset{a}{\ket{\mathbf{x}}}\overset{b}{\ket{\mathbf{x}}}\right)
\end{split}
\end{equation*}
where Alice's qubits are labelled with an ``$a$'' and Bob's are
labelled with a ``$b$'', and $\mbox{H}(\mathbf{x})$ is the number
of $1$s in the string $\mathbf{x}$, also known as the Hamming
weight of $\mathbf{x}$. On the right hand side of the equality,
the state is written in terms of the symmetric basis. The
symmetric space is an $(n+1)$-dimensional subspace of the
$2^n$-dimensional state-space for the register. The $i^{th}$
symmetric basis state is a uniform superposition of the
computational basis states having Hamming weight $i$.  Alice and
Bob can each measure the Hamming weight of their half of the state
$\ket{\Psi}^n$. The measurement is implemented by introducing an
ancilla of size $O(\log n)$.  A sequence of controlled-[add 1]
operations is used to add the Hamming weight of each qubit of
$\ket{\Psi}$ into the ancilla. This is implemented by network
shown in Figure \ref{hwfig}.

Suppose Alice measures the Hamming weight of $\ket{\Psi}$ and
obtains the result $\ket{j}$ (Bob will measure the same $j$
whenever he performs the same measurement). This state after the
measurement is
\begin{equation*}
\frac{1}{\sqrt{\binom{n}{j}}}\sum_{\mbox{H}(\mathbf{x}) = j}
\overset{a}{\ket{\mathbf{x}}}\overset{b}{\ket{\mathbf{x}}}
\end{equation*}
which can be thought of as a superposition of $\binom{n}{j}$
$n$-bit strings. (Of course, the measurement is not necessary, and
the remainder of the algorithm could be controlled quantumly upon
the value $j$.) Let $r = \lceil\log_2 \binom{n}{j}\rceil$. Define
a function $f$ on these $\binom{n}{j}$ strings that maps the
${n\choose j}$ strings of length $n$ with Hamming weight $j$ (in
lexicographic order) to the integers from $0$ to ${n \choose
j}-1$:
\begin{equation*}
\begin{split}
f(00\ldots00\underset{j}{\underbrace{11\ldots1}}) & = 00\ldots0
\\ f(00\ldots10\underset{j-1}{\underbrace{11\ldots1}}) & = 00\ldots1
\\
\hspace{1cm}\vdots &
\\
f(\underset{j}{\underbrace{11\ldots1}}00\ldots0) & =
\binom{n}{j}-1 = m
\\ & =
\underset{n-r}{\underbrace{00\ldots0}}\underset{r}{\underbrace{m}}.
\end{split}
\end{equation*}
We can extend $f$  so that it defines a permutation of \emph{all}
$n$-bit strings.  Then we have:
\begin{equation*}
\begin{split}
\frac{1}{\sqrt{\binom{n}{j}}}\sum_{\mbox{H}(x)=j}\ket{\mathbf{x}}\ket{\mathbf{x}}
\overset{f}{\longrightarrow} &
\sum_{\mbox{H}(x)=j}\ket{f(\mathbf{x})}
\ket{f(\mathbf{x})} \\
 & = \sum_{y=0}^{\binom{n}{j}-1}
 \underset{n-r}{\underbrace{\ket{\mathbf{0}}}}
 \underset{r}{\underbrace{\ket{\mathbf{y}}}}
 \ket{\mathbf{0}}\ket{\mathbf{y}}.
\end{split}
\end{equation*}
If $\binom{n}{j} = 2^r$, then ignoring the first $n-k$ bits on
both sides gives us
\begin{equation*}
\sum_{\mathbf{y}=0}^{2^r-1}\ket{\mathbf{y}}\ket{\mathbf{y}}
\end{equation*}
which is $r$ EPR-pairs, and the entanglement of $\ket{\Psi}$ has
been concentrated.

However, in general,  $\binom{n}{j}$ will not be a power of $2$.
Let $k=\lfloor \log_2\binom{n}{j}\rfloor+1$. We describe a quantum
network that will produce some number $0\leq l \leq k-1$ of
EPR-pairs (we use this definition for $k$ in place of the previous
definition for $r$ for convenience in describing a network that
will behave the same whether or not $\binom{n}{j}$ is a power of
2). The expected number of EPR pairs will be at least $k-2$. We
illustrate this for $n=3$ entangled pairs of qubits. Consider the
binary representation $\binom{n}{j} = x_2x_1x_0 =
x_2\cdot2^2+x_1\cdot2^1+x_0\cdot2^0$. We have
\begin{equation} \label{bigsum}
\sum_{\mathbf{y}=0}^{\binom{n}{j}-1}\ket{\mathbf{y}}\ket{\mathbf{y}}
=\sum_{\mathbf{y}=0}^{x_2\cdot2^2-1}\ket{\mathbf{y}}\ket{\mathbf{y}}
+ \sum_{\mathbf{y}=x_2\cdot2^2}^{x_2\cdot2^2+x_1\cdot2^1-1}
\ket{\mathbf{y}}\ket{\mathbf{y}} +
\sum_{\mathbf{y}=x_2\cdot2^2+x_1\cdot2^1}^{x_2\cdot2^2+x_1\cdot2^1+x_0\cdot2^0-1}
\ket{\mathbf{y}}\ket{\mathbf{y}}.
\end{equation}
Notice that if $x_2=1$ then the above sum includes
$000\leq\mathbf{y}\leq011$.  These are included in
$\sum_{\mathbf{y}=0}^{x_2\cdot2^2-1}\ket{\mathbf{y}}\ket{\mathbf{y}}$
which is the first term on the right side of (\ref{bigsum}) (if
$x_2=0$, then this term is empty). Similarly, if $x_1 = 1$ the sum
includes $x_200\leq\mathbf{y}\leq x_201$ and if $x_0 = 1$ it
includes $x_2x_10\leq\mathbf{y}\leq x_2x_11$.  So we can write the
sum (\ref{bigsum}) as follows:
\begin{equation} \label{sum2}
\sum_{\mathbf{y}=0}^{\binom{n}{j}-1}\ket{\mathbf{y}}\ket{\mathbf{y}}
= x_2 \sum_{\mathbf{y} = 000}^{011}
\ket{\mathbf{y}}\ket{\mathbf{y}} + x_1 \sum_{\mathbf{y} =
x_200}^{x_201} \ket{\mathbf{y}}\ket{\mathbf{y}} +
x_0\sum_{\mathbf{y} = x_2x_10}^{x_2x_10}
\ket{\mathbf{y}}\ket{\mathbf{y}}.
\end{equation}
In other words, for each $j$ such that $x_j=1$, we have the
superposition of $2^j$ strings. Alice and Bob wish to project to
one of these superpositions of $2^j$ strings, since that will
provide them with $j$ EPR pairs.

The first term on the right side of (\ref{sum2}) contains the
strings $000$, $001$, $010$, $011$; all the strings beginning with
a $0$. Suppose Alice performs a measurement of the qubit in the
leftmost position (i.e. corresponding to $y_2$) of her share of
the state
$\sum_{\mathbf{y}=0}^{\binom{n}{j}-1}\ket{\mathbf{y}}\ket{\mathbf{y}}$
(Bob will obtain the same result whenever he performs the
analogous measurement on his share).  In addition, Alice also has
the corresponding bit $x_2$ in a register containing the binary
expansion of $\binom{n}{j}$. There are three cases to consider:
\begin{enumerate}
\item[CASE 1:]
$y_2=0$ and $x_2=1$.

In this case the joint state after the measurement is
\[ \sum_{\mathbf{y}=000}^{011} \ket{\mathbf{y}}\ket{\mathbf{y}} \]
which is the first term on the right side of (\ref{sum2}).
Ignoring $\ket{y_2}$, this is 2 EPR pairs.
\item[CASE 2:]
$y_2=0$ and $x_2=0$.

 The state after the measurement is
\[x_1 \sum_{\mathbf{y} =
000}^{001} \ket{\mathbf{y}}\ket{\mathbf{y}} + x_0\sum_{\mathbf{y}
= 0x_10}^{0x_10} \ket{\mathbf{y}}\ket{\mathbf{y}}.\] Ignoring the
leftmost qubit $\ket{y_2}$ this is equal to \[ x_1
\sum_{\mathbf{y} = 00}^{01} \ket{\mathbf{y}}\ket{\mathbf{y}} +
x_0\sum_{\mathbf{y} = x_10}^{x_10}
\ket{\mathbf{y}}\ket{\mathbf{y}}. \]

\item[CASE 3:]
$y_2=1$.

In this case we know $x_2=1$.  So the post-measurement state is
\[x_1 \sum_{\mathbf{y} = 100}^{101}
\ket{\mathbf{y}}\ket{\mathbf{y}} + x_0\sum_{\mathbf{y} =
1x_10}^{1x_10} \ket{\mathbf{y}}\ket{\mathbf{y}}.\] Ignoring the
leftmost qubit $\ket{y_2}$ this is equal to \[x_1 \sum_{\mathbf{y}
= 00}^{01} \ket{\mathbf{y}}\ket{\mathbf{y}} + x_0\sum_{\mathbf{y}
= x_10}^{x_10} \ket{\mathbf{y}}\ket{\mathbf{y}}.\]
\end{enumerate}

If Alice's measurement results in case 1, then the entanglement
has been concentrated, and she makes no further measurements.
Cases 2 and 3 both leave Alice and Bob with the state $x_1
\sum_{\mathbf{y} = 00}^{01} \ket{\mathbf{y}}\ket{\mathbf{y}} +
x_0\sum_{\mathbf{y} = x_10}^{x_10}
\ket{\mathbf{y}}\ket{\mathbf{y}}$.  In either of these cases,
Alice discards the leftmost qubit $\ket{y_2}$.  She then repeats
the measurement procedure, where this time the leftmost bits being
measured are $y_1$ and $x_1$. The analogous three cases are
considered again.

This time case 1 would result in the post measurement state
$\sum_{\mathbf{y} = 00}^{01} \ket{\mathbf{y}}\ket{\mathbf{y}}$.
Ignoring the leftmost qubit $y_1$, this gives 1 EPR pair, and the
procedure stops.  Cases 2 and 3 both result in the post
measurement state
$x_0\sum_{\mathbf{y}=0}^0\ket{\mathbf{y}}\ket{\mathbf{y}}$, giving
0 EPR pairs.

It is easy to generalise this approach for $k = \lfloor\log_2
\binom{n}{j}\rfloor+1$.  Alice (or Bob) measures (locally) the
qubits $y_{k-1},\ldots,y_1$, from ``left-to-right'', at each step
checking the value of the corresponding bit $x_i$ in the binary
expansion of $\binom{n}{j}$.  She does this until, at some
iteration $l$ (where the first iteration is indexed 0), she finds
$\ket{y_{k-1-l}}=\ket{0}$ and the corresponding bit $x_{k-1-l}=1$.
When this occurs the procedure stops, having distilled $k-l-1$ EPR
pairs.

\begin{figure}[h]
\begin{center}
\epsfig{file=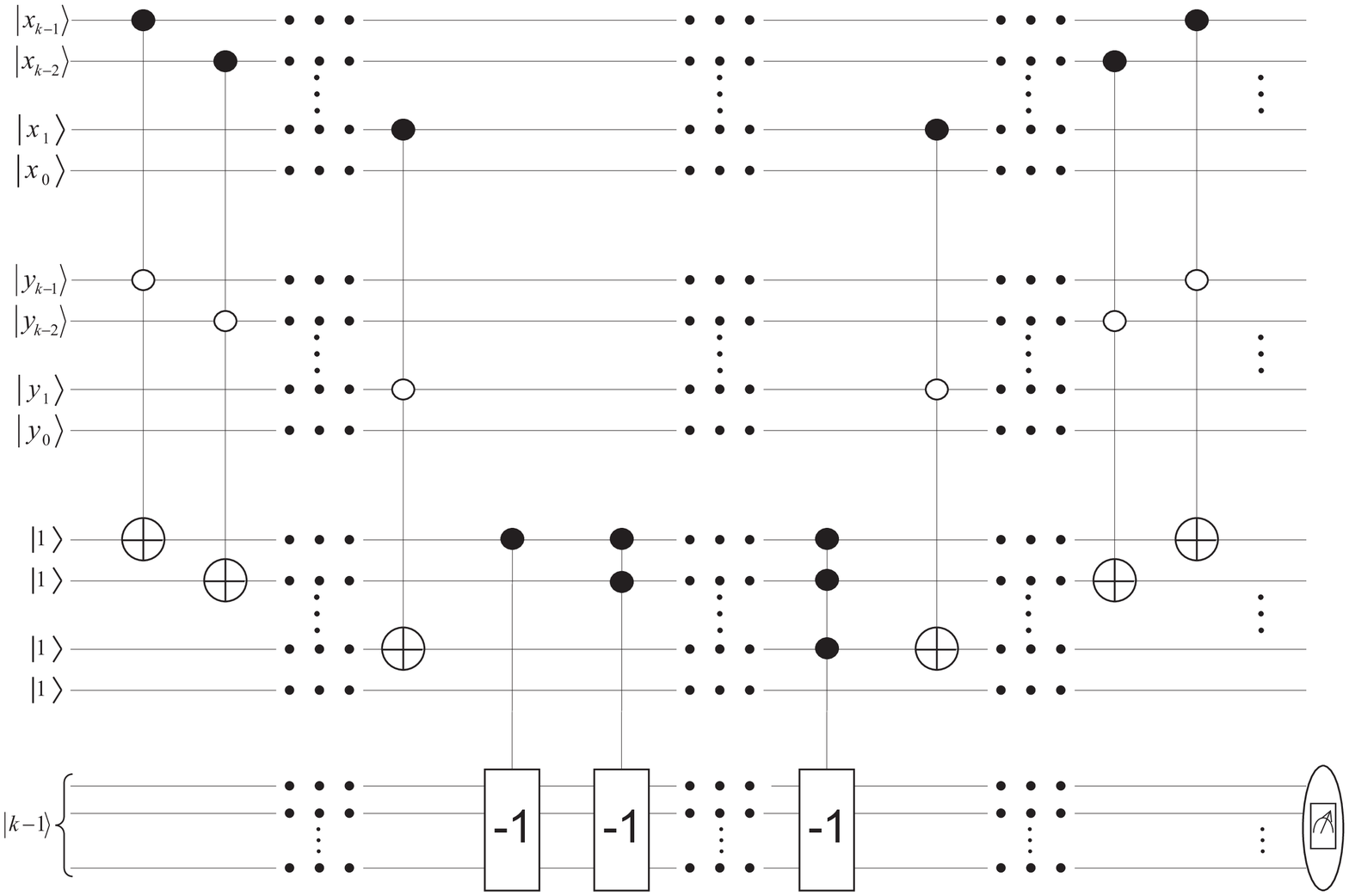,height=8cm,width=10cm} \caption{Network
to measure how many EPR pairs have been distilled.}\label{netfig}
\end{center}
\end{figure}

A quantum network implementing the procedure is shown in Figure
\ref{netfig}.  Since the Hamming weight $j$ has been measured,
$\binom{n}{j}$ can be efficiently computed.  The binary
representation of $\binom{n}{j}$ is encoded in a register
$\ket{x_{k-1}\ldots x_0}$. The network makes use of an ancilla of
size $k$, initially in the state $\ket{1}^k$. We refer to this
ancilla as the ``control ancilla'', and label its qubits of the by
$\ket{t_i}$ for $0\leq i\leq k-1$. For each $i$, $\ket{t_i}$ is
switched to $\ket{0}$ if both $\ket{y_i}=\ket{0}$ and
$\ket{x_i}=\ket{1}$. This is achieved using a sequence of doubly
controlled NOT gates in the first stage of the network, where the
NOT is applied to the target qubit if the first control qubit is
in state $\ket{1}$ and the second control qubit is in state
$\ket{0}$. Another ancilla of size $O(\log(k-1))$, which we will
call the ``measurement ancilla'' is initially in the state
$\ket{k-1}$. In the second stage of the network, the value of the
measurement ancilla is decremented by a sequence of
controlled-[subtract 1] gates, controlled successively on each of
the $\ket{t_i}$ in the control ancilla.  The net effect of the
first two stages of the network is to decrement the measurement
ancilla by one for each pair $(\ket{x_i},\ket{y_i})$ until one
such pair is found with $(\ket{x_i}=\ket{1},\ket{y_i}=\ket{0})$.
After such a pair is encountered, the measurement ancilla is not
decremented any more. In order to reverse the effect of any
coupling that the network may have introduced between the primary
register $\ket{\mathbf{y}}$ and the ancilla $\ket{\mathbf{t}}$,
the same sequence of doubly controlled NOT gates that was used in
the first stage of the network is applied again in the third
stage.

The control ancilla has been reset to its initial state by the
third stage of the network, and the register $\ket{\mathbf{x}}$
containing the binary expansion of $\binom{n}{j}$ is in a fixed
computational basis state, since the value of $j$ was fixed by the
Hamming weight measurement performed earlier.  Ignoring the state
the control ancilla and the register $\ket{\bf{x}}$, the joint
state of Alice's system $\ket{\mathbf{y}}$ and the measurement
ancilla just before the final measurement is
\[ x_{k-1}\sum_{y=0^k}^{01^{k-1}}\ket{\mathbf{y}}\ket{k-1}+
x_{k-2}\sum_{y=x_{k-1}0^{k-1}}^{x_{k-1}01^{k-2}}\ket{\mathbf{y}}\ket{k-2}+\cdots
+x_0\sum_{y=x_{k-1}\cdots x_10}^{y=x_{k-1}\cdots
x_10}\ket{\mathbf{y}}\ket{0}. \]

The string $x_{k-1} x_{k-2} \ldots x_{k-l}$ correspond to the the
leftmost $l$ bits in the binary representation of $\binom{n}{j}$.
After the measurement of the control ancilla in the computational
basis, the state is
\[ \sum_{y=x_{k-1} x_{k-2} \ldots x_{k-l}0^{k-l}}^{x_{k-1} x_{k-2} \ldots x_{k-l}01^{k-l-1}}\ket{\mathbf{y}}\ket{k-l-1} \]
for some $0\leq l \leq k-1$.  Ignoring the leftmost $l+1$ qubits,
this is
\[ \sum_{y=0^{k-l-1}}^{1^{k-l-1}}\ket{\mathbf{y}}\ket{k-l-1}. \]
Each ignoring their respective leftmost $l+1$ qubits, the joint
Alice-Bob state is
\[ \sum_{y=0^{k-l-1}}^{1^{k-l-1}}\ket{\mathbf{y}}\ket{\mathbf{y}} \]
which is $k-l-1$ EPR pairs. Note that the state of Alice's ancilla
(and Bob's, if he performs the same measurement procedure on his
share of the state) indicates the number of EPR pairs that have
been distilled.

It should be noted that Alice and Bob can each carry out the above
procedure locally, and they will obtain the same results.
Alternatively, Alice could perform the Hamming weight computation
locally and send the result to Bob.  Alice and Bob would both
perform the permutation $f$ by the method detailed in section
\ref{network}). In the last stage of the procedure, either Alice
or Bob could perform the computation to determine the number of
EPR pairs that have been distilled, and send the result to the
other.

It can be shown that given the superposition
\begin{equation*}
\sum_{\mathbf{y}=0}^{\binom{n}{j}-1}\ket{\mathbf{y}}\ket{\mathbf{y}}
\end{equation*}
the average number of
EPR-pairs produced using this approach is:
\begin{equation*}
\sum_{i=1}^{\lfloor\log_2{\binom{n}{j}}\rfloor}x_{n-i}\frac{2^i}{\binom{n}{j}}
\geq \lfloor\log_2{\binom{n}{j}}\rfloor-1=k-2.
\end{equation*}

\section{Implementing the permutation $f$}\label{network}
The key step in the entanglement concentration protocol is the
permutation $f$.  We need to know how to implement this function.
Recall that we start with a superposition of $\binom{n}{j}$
strings $\mathbf{x}$, each having $j$ $1s$ and $n-j$ $0s$.  We
want $f$ to impose a lexicographic ordering on these strings.

Consider the following:
\begin{equation*}
\begin{split}
0\hspace{1mm}0\hspace{1mm}\ldots\hspace{1mm}0\hspace{3mm} &
\overset{f^{-1}}{\longrightarrow}
\hspace{3mm}0\hspace{1mm}0\hspace{1mm}\ldots\hspace{1mm}0
\hspace{1mm}0\hspace{1mm}\underset{j}{\underbrace{1\hspace{1mm}1\hspace{1mm}\ldots\hspace{1mm}1}}
\\0\hspace{1mm}0\hspace{1mm}\ldots\hspace{1mm}1\hspace{3mm} &
\overset{f^{-1}}{\longrightarrow}\hspace{3mm}0\hspace{1mm}0\hspace{1mm}\ldots\hspace{1mm}
1\hspace{1mm}0\hspace{1mm}\underset{j-1}{\underbrace
{1\hspace{1mm}1\hspace{1mm}\ldots\hspace{1mm}1}}
\\
& \hspace{0.5cm}\vdots
\\
\binom{n-1}{j}-1 \hspace{3mm} & \overset{f^{-1}}{\longrightarrow}\hspace{3mm}
0\hspace{1mm}\underset{j}{\underbrace{1\hspace{1mm}1\hspace{1mm}\ldots\hspace{1mm}1}}
\hspace{1mm}0\hspace{1mm}0\hspace{1mm}\ldots\hspace{1mm}0
\\
\binom{n-1}{j} \hspace{3mm}& \overset{f^{-1}}{\longrightarrow}\hspace{3mm}
1\hspace{1mm}0\hspace{1mm}\ldots\hspace{1mm}0\hspace{1mm}
\underset{j-1}{\underbrace{1\hspace{1mm}1\hspace{1mm}\ldots\hspace{1mm}1}}
\\
& \hspace{0.5cm}\vdots
\\
\binom{n}{j}-1 \hspace{3mm}&
\overset{f^{-1}}{\longrightarrow}\hspace{3mm}
\underset{j}{\underbrace{1\hspace{1mm}1\hspace{1mm}\ldots\hspace{1mm}1}}
\hspace{1mm}0\hspace{1mm}0\hspace{1mm}\ldots\hspace{1mm}0.
\end{split}
\end{equation*}

The first $\binom{n-1}{j}$ strings have a $0$ in the first bit
position, and the remaining strings have a $1$ in the first bit
position.  Define $[\mathbf{y}]_{n,j} $ to be the $\mathbf{y}^{th}$
largest string (treating the string as an integer represented in
binary) of length $n$ with Hamming weight $j$. Using this notation,
the method for implementing $f$ is captured by the following
recurrence:
\begin{equation} \label{recur}
\begin{split}
[\mathbf{y}]_{n,j} & = 0[\mathbf{y}]_{n-1,j} \hspace{2.8cm} \mbox{
if }
 0\leq \mathbf{y} <
\binom{n-1}{j}
\\& = 1\left[\mathbf{y}-\binom{n-1}{j}\right]_{n-1,j-1} \mbox{ if }
\binom{n-1}{j} \leq \mathbf{y} < \binom{n}{j}.
\end{split}
\end{equation}
We describe how the permutation $f^{-1}$ can be implemented on a
quantum computer. Let $[\mathbf{y}]_{n,j} = b_1b_2\ldots b_n$.
Then start with a string between $00\ldots0$ and $\binom{n}{j}-1$,
and ancilla holding the values $n$, $j$, and a space for the
output strings $f^{-1}(\mathbf{y})$:
\begin{equation*}
\ket{\mathbf{y}}\ket{n}\ket{j}\ket{00\ldots0}.
\end{equation*}
Apply an operator $T$ which performs the following mapping:
\begin{equation*}
\begin{split}
\ket{\mathbf{y}}\ket{n}\ket{j}\ket{00\ldots0} & \overset{T}{\longrightarrow}
\ket{\mathbf{y}}\ket{n}\ket{j}\ket{00\ldots0} \hspace{5mm}
\mbox{ if }\hspace{2mm} 0 \leq \mathbf{y} < \binom{n-1}{j}
\\ & \overset{T}{\longrightarrow}
\ket{\mathbf{y}}\ket{n}\ket{j}\ket{10\ldots0} \hspace{5mm} \mbox{
if }\hspace{2mm} \binom{n-1}{j} \leq \mathbf{y} < \binom{n}{j}.
\end{split}
\end{equation*}
The result is:
\begin{equation*}
\ket{\mathbf{y}}\ket{n}\ket{j}\ket{00\ldots0} \overset{T}{\longrightarrow}
\ket{\mathbf{y}}\ket{n}\ket{j}\ket{b_10\ldots0}
\end{equation*}
for some $b_1 \in \{0,1\}$.  Then perform the following
subtraction operation $S$, controlled quantumly on the value of
$b_1$:
\begin{equation*}
\begin{split}
\ket{\mathbf{y}}\ket{n}\ket{j} & \overset{S}{\longrightarrow}
\ket{\mathbf{y}}\ket{n-1}\ket{j} \hspace{33mm}
\mbox{ if }\hspace{2mm} b_1 = 0
\\ & \overset{S}{\longrightarrow}
\ket{\mathbf{y}-\binom{n-1}{j}}\ket{n-1}\ket{j-1} \hspace{5mm}
\mbox{ if }\hspace{2mm} b_1 = 1.
\end{split}
\end{equation*}
Then repeat $T$ and $S$, this time on only the right-most $n-1$
bits of the registers $\ket{y}$ and $\ket{b_10..0}$.  Applying $T$
and $S$ in this way, a total of $n$ times, realises the recurrence
(\ref{recur}), and gives us an implementation of $f^{-1}$:
\begin{equation*}
\ket{\mathbf{y}}
\overset{f^{-1}}{\longrightarrow}\ket{[{\mathbf{y}}]_{n,j}}
\end{equation*}
and thus the same network maps
\begin{equation*}
\sum_{\mathbf{y}=\mathbf{0}}^{\binom{n}{j}-1}\ket{\mathbf{y}}
\overset{f^{-1}}{\longrightarrow}\sum_{H(\mathbf{x}) =
j}\ket{\mathbf{x}}.
\end{equation*}
The permutation $f$ is realised simply by running this procedure
backwards.

\section{An algorithm for entanglement concentration}\label{summary}
We now have the tools to state an algorithm for
implementing the entanglement concentration protocol described in
section 4.1. The algorithm is the following:
\begin{enumerate}
\item Begin with the state $\ket{\Psi} = (\alpha\ket{00}+\beta\ket{11})^n$.
\item \label{step2}
Alice and Bob each perform a Hamming-weight measurement on their
half of $\ket{\Psi}$, obtaining the same result $\ket{j}$.
\item Alice and Bob each perform the permutation $f$ on the resulting superposition.
\item Alice and Bob each use the network of Figure \ref{netfig} to determine how many EPR pairs they share.
\item The result is some known number of perfect EPR pairs.  For a particular $j$, the expected number is between
$k-2$ and $k-1$ where \mbox{$k = \lfloor\log_2 {n \choose
j}\rfloor+1$}.
\end{enumerate}
Each of the above steps have been detailed in the preceding
sections, and so we have a complete description of the
implementation.  Since the probability of measuring $\ket{j}$ in
step \ref{step2} is \[ |\alpha^2|^{n-j} \left(1 - |\alpha^2|
\right)^j {n \choose j}, \] the expected number of EPR pairs is at
least
\begin{eqnarray*}
\sum_{j=0}^{n} |\alpha^2|^{n-j} \left(1 - |\alpha^2| \right)^j {n
\choose j} ( \lfloor\log_2 {n \choose j}\rfloor-1)
\end{eqnarray*}
and comparing to equation (\ref{equation1}) shows that we get at
least
\[ n H(|\alpha^2|)
- O(\log n) \] EPR pairs on average.  Note that the theoretical
maximum is $nH(|\alpha^2|)$.

\end{document}